\documentclass[aps,prc,twocolumn,showpacs,twoside]{revtex4}
\usepackage{graphicx,bm}

\begin{document}

\newcommand{\tlab}{T_{\mbox{\scriptsize lab}}}
\newcommand{\plab}{p_{\mbox{\scriptsize lab}}}
\newcommand{\pppi}{\bar p p \rightarrow \pi^-\pi^+}
\newcommand{\ppkk}{\bar p p \rightarrow K^-K^+}
\newcommand{\s}{^3S_1}
\newcommand{\p}{^3\!P_0}
\newcommand{\ana}{A_{0n}}
\newcommand{\dsig}{d\sigma/d\Omega}
\newcommand{\pvec}{{\mathbf p}} 
\newcommand{\rvec}{{\mathbf r}}
\newcommand{\cm}{{\mathrm{cm}}}
\newcommand{\dvec}{\mathrm{d}{\mathbf p}}

\title{Relativistic effects and angular dependence in the reaction $\pppi$}
\author{B.~El-Bennich}
\email{bennich@physics.rutgers.edu}
\author{W.M.~Kloet}
\affiliation{Department of Physics and Astronomy, Rutgers University, \\ 
             136 Frelinghuysen Road, Piscataway, New Jersey 08854-8019, USA}
\date{\today}

\begin{abstract}
We present a new fit to the LEAR data on $\pppi$ differential cross sections and analyzing powers motivated by relativistic 
considerations. Within a quark model describing this annihilation we argue, since the pions are highly energetic, that 
relativistic effects cannot be neglected. The intrinsic pion wave functions are Lorentz transformed to the center of mass frame. 
This change in quark geometry gives rise to additional angular dependence in the transition operators and results in a relative 
enhancement of higher $J\ge 2$ partial wave amplitudes. The fit to the data is improved significantly.
\end{abstract}

\maketitle

\section{Introduction \label{intro}}

We recently studied effects of particle size and final-state interactions in the reaction $\pppi$ within the framework of a constituent 
quark model \cite{elbennich1}. The aim was to improve previous attempts \cite{kloet1,kloet2} to describe the LEAR data on $\pppi$ 
differential cross sections  $d\sigma/d\Omega$ and analyzing powers $A_{0n}$ in the momentum range from 360 to 1550 MeV/$c$ 
\cite{hasan}. As of yet, theoretical approaches, whether using a baryonic or a quark picture~\cite{kloet2,moussallam83,mull91,mull92,muhm96,yan96},
have not been successful in reproducing the characteristic double-dip structure of the $A_{0n}$ observables nor the $d\sigma/d\Omega$ forward 
peaks at low momenta. The large variations of the LEAR observables as a function of the angle portends the presence of a substantial number 
of partial wave amplitudes already at low energies. In fact, the experimental data on differential cross sections as well as 
those on asymmetries point to a significant $J=2$, $J=3$ and even higher $J$ contributions~\cite{oakden94,hasan94,kloet96,martin97}. 
Model calculations \cite{kloet2,moussallam83,mull91,mull92,muhm96,yan96}, however, lead invariably to scattering amplitudes which 
are strongly dominated by total angular momentum $J=0$ and $J=1$. This is due to a rather short range of the annihilation mechanism in 
these models. 

In order to study possible higher partial wave $J\ge 2$ contributions, the role of final-state $\pi\pi$ interactions has been investigated 
in Ref.~\cite{elbennich1,mull91,mull92,greenfinal}. For example, in Ref.~\cite{elbennich1} we made use of the non-planar $R2$ quark-flow diagrams, 
in which a $\bar qq$ pair in either a $\s$ or a $\p$ state is annihilated and momentum is transferred to a remaining quark or antiquark as 
discussed in Ref.~\cite{kloet1,kloet2}.
Switching on the $\pi\pi$ interaction moderately improved the fit of $A_{0n}$, in particular the double-dip structure at low 
momenta is slightly more pronounced. This is caused by a readjustment of the strengths of the helicity amplitudes of different total
angular momentum $J$. Nonetheless, the predictions for $d\sigma/d\Omega$ showed only a modest improvement over the model
without final-state interactions. 

The main improvement obtained in Ref.~\cite{elbennich1}, however, is due to a different aspect, namely a readjustment of the size parameters 
of the intrinsic proton and pion wave functions, so that the radii of the proton, antiproton, and pions increase. In a final fit these radii are 
larger by about 7\% than the respective measured charge distribution radii. This runs contrary to the view that the valence quarks occupy a smaller 
volume than indicated by the charge radii. Nevertheless, introducing larger radii as well as final-state interactions, improves the quality of the 
fit in Ref.~\cite{elbennich1} dramatically -- the forward and backward peaks of $d\sigma/d\Omega$ are very well reproduced and so are the characteristic 
double-dip structures of $A_{0n}$. It should come as no surprise that the relative contribution of the higher $J\ge 2$ amplitudes in Ref.~\cite{elbennich1} 
turns out to be significantly larger than in Ref.~\cite{kloet2} 

In this paper, we report an alternative approach that also leads to enhancement of the higher partial waves. It addresses the relativistic effects due 
to Lorentz transformed intrinsic pion wave functions in the reaction $\pppi$. The reason for this rather different approach stems from the observation
that at the center-of-mass (c.m.) energies $\sqrt{s}$ considered in the LEAR experiment $\pppi$, the produced pions are highly energetic. 
The relativistic factor in this energy range is $\gamma=E_\cm/2m_{\pi} \simeq 6.8-8.0$, which means the pions are ejected at speeds $v\simeq 0.98c$. 
One therefore expects considerable relativistic effects due to their modified internal structure. In the c.m. frame in which the transition amplitudes 
are calculated, the intrinsic pion wave functions employed must be Lorentz transformed from the pion rest frame.
In our model \cite{elbennich1,kloet1,kloet2} the intrinsic pion as well as the proton and antiproton wave functions are of Gaussian form in their 
respective rest frames. We neglect the distortions of the proton and antiproton intrinsic wave function due to their much smaller kinetic energy 
and larger mass.

The relativistic effects have several consequences. First, the intrinsic geometry of the pions is modified --- instead of spherical particles one
deals now with highly flattened ellipsoids in the c.m. frame. Obviously, the reaction geometry is altered also since the boosted pion wave functions 
enter the computation of the transition operators, which are obtained from an overlap integral of initial antiproton-proton, final pion-pion wave 
functions and the $\s$ or $\p$ annihilation mechanism. Details of this calculation were presented in Ref.~\cite{elbennich2}. Secondly, due to less 
overlap of the pion and antiproton-proton intrinsic wave functions, the annihilation range {\em actually\/} decreases but this effect is far from 
isotropic. As a result the transition operators exhibit significant additional angle dependence. Finally, the annihilation amplitudes, already 
non-local in the case of non-relativistic wave functions, become explicitly dependent on the c.m. energy $\sqrt{s}$ via the boost factor 
$\gamma$.

In Ref.~\cite{elbennich2} it was shown that the relativistic transformation of the spatial part of the intrinsic pion wave functions 
introduces new angle dependent terms in the transition operators. These new terms also depend on the boost factor $\gamma$. In this paper, we present a new fit to the LEAR data on 
$\pppi$ using the $R2$ transition operators of Ref.~\cite{elbennich2}, which supports our claim that relativistic considerations lead
to a strongly modified angular momentum content of the helicity amplitudes for $0\le J\le 4$. Especially the amplitudes for $J\ge 2$ are 
amplified considerably. Even though we keep the original particle radii as in \cite{kloet1,kloet2}, i.e. smaller than the 
corresponding charge distribution radii, we achieve a very good reproduction of the LEAR observables $d\sigma/d\Omega$ and $A_{0n}$,
comparable to the one in Ref.~\cite{elbennich1}.

\section{Relativistic modifications \label{two}}

We summarize the effects of Lorentz transforming intrinsic pion wave functions on $\pppi$ annihilation operators within the constituent quark model.
The actual details may be found in Ref.~\cite{elbennich2}. In the pion rest frame, the radial part of its intrinsic wave 
function is described by a Gaussian of the form  
\begin{equation}
 \psi_\pi(\rvec_i,\rvec_j) = N_\pi \exp \left \{ -\frac{\beta}{2}
 \sum_{\alpha=i,j} (\rvec_\alpha - \mathbf{R}_{\!\pi} )^2 \right  \} ,
\label{pionnoboost}
\end{equation}
which reads in the c.m. frame
\begin{eqnarray}
\lefteqn{\hspace*{-3mm}\psi_\pi(l^{-1}\rvec_i,l^{-1}\rvec_j) =} \nonumber \\
 & \hspace*{-4mm} = &\hspace*{-2mm} \tilde N_\pi \exp \left \{\! -\frac{\beta}{2} \sum_{\alpha=i,j}\! 
   \left [ (\rvec_\alpha-\mathbf{R}_{\!\pi} )_\perp^2\! 
  +\!\gamma^2(\rvec_{\alpha}-\mathbf{R}_{\!\pi} )^2_\|\right ] \!\right \}\!\!. 
\label{rwaveboost}
\end{eqnarray}
The vectors $\rvec_i$ and $\rvec_j$ are the quark and antiquark coordinates, $\mathbf{R}_{\!\pi}=\frac{1}{2}(\rvec_i+\rvec_j)$ 
is the pion coordinate, and $\beta$ is the size parameter. 
As mentioned above, in previous work~\cite{elbennich1} we varied the value of $\beta$ in order to obtain an increase in the annihilation range. 
In this paper we use throughout the fixed value $\beta=3.23~$fm$^{-2}$, which corresponds to a pion radius of 0.48~fm \cite{kloet1,kloet2,green} . 
The new normalization factor $\tilde N_\pi=\sqrt{\gamma}\,N_\pi$ is due to the condition that $\psi_\pi$ be normalized to unity in the c.m.

The c.m. wave functions of Eq.~(\ref{rwaveboost}) are used to compute the transition operators for the $\p$ and $\s$ annihilation amplitudes. 
It is shown in Ref.~\cite{elbennich2} that these operators acquire new terms if compared with the non-relativistic expressions \cite{kloet1} 
and which manifestly introduce additional angular dependence. We here briefly recapitulate the results for the $R2$ diagrams from \cite{elbennich2}. 
The complete form of the $R2$ transition operators for the {\em vacuum\/} $\p$ amplitude is
\begin{widetext}
\begin{eqnarray}
\hat T (\p) = i\mathcal{N}\Big [ A_V \bm{\sigma}\!\cdot\!\mathbf{R'}\sinh(C\,\mathbf{R}\!\cdot\!\mathbf{R'}) 
 + B_V \bm{\sigma}\!\cdot\!\mathbf{R}\cosh(C\,\mathbf{R}\!\cdot\!\mathbf{R'}) 
 + C_V (\bm{\sigma}\!\cdot\!\mathbf{\hat R'})\,R\cos\theta \cosh(C\,\mathbf{R}\!\cdot\!\mathbf{R'})\Big ] \times \nonumber \\
   \times\,  \exp \{A\mathbf{R'}^2+B\mathbf{R}^2+D \mathbf{R}^2\cos^2\theta \} .
\label{tot1} 
\end{eqnarray}
The total $\s$ amplitude for the {\em longitudinal\/} component is given by
\begin{eqnarray}
\hat T (\s^L) = i\mathcal{N}\Big [ A_L \bm{\sigma}\!\cdot\!\mathbf{R'}\sinh(C\,\mathbf{R}\!\cdot\!\mathbf{R'}) 
 + B_L \bm{\sigma}\!\cdot\!\mathbf{R}\cosh(C\,\mathbf{R}\!\cdot\!\mathbf{R'}
 + C_L (\bm{\sigma}\!\cdot\!\mathbf{\hat R'})\,R\cos\theta \cosh(C\,\mathbf{R}\!\cdot\!\mathbf{R'})\Big ] \times \nonumber \\
 \times\, \exp \{A\mathbf{R'}^2+B\mathbf{R}^2+D \mathbf{R}^2\cos^2\theta \} 
\label{tot2}
\end{eqnarray}
and
\begin{eqnarray}
 \hat T (\s^T)=\mathcal{N}\Big [ A_T \bm{\sigma}\!\cdot\!\mathbf{R'}\cosh(C\,\mathbf{R}\!\cdot\!\mathbf{R'}) 
  + B_T \bm{\sigma}\!\cdot\!\mathbf{R}\sinh(C\,\mathbf{R}\!\cdot\!\mathbf{R'}) +
    C_T (\bm{\sigma}\!\cdot\!\mathbf{\hat R'})\,R\cos\theta \sinh(C\,\mathbf{R}\!\cdot\!\mathbf{R'})\Big ] \times \nonumber \\
   \times\, \exp \{A\mathbf{R'}^2+B\mathbf{R}^2+D \mathbf{R}^2\cos^2\theta \} 
\label{tot3}
\end{eqnarray}
\end{widetext}
\begin{figure*}[t]
\begin{center}
\includegraphics[angle=90,scale=0.55]{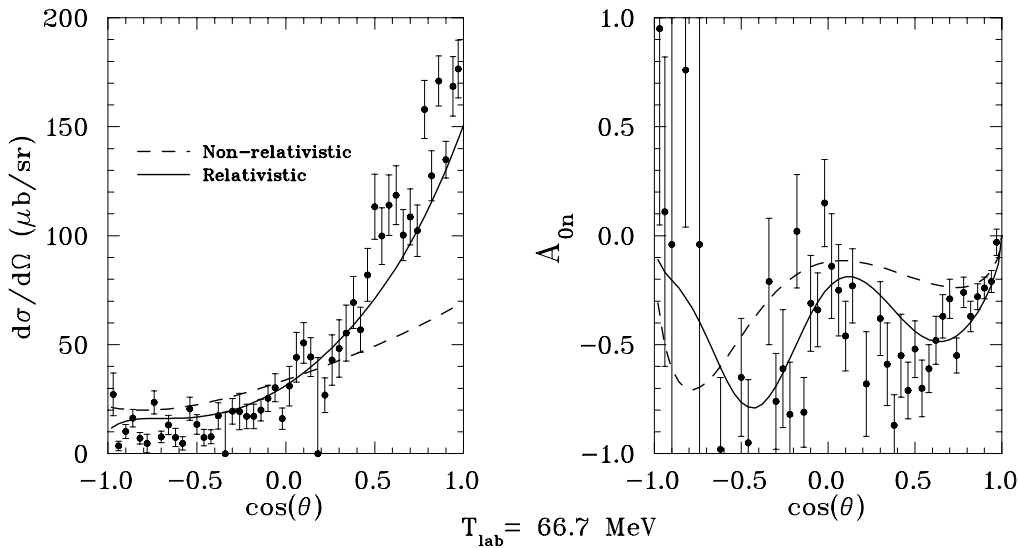}
\end{center}
\caption{\small Differential cross section and analyzing power of the reaction $\pppi$ at $\tlab$= 66.7 MeV ($\plab$= 360 MeV/c).
         Experimental data are from Hasan {\em et al}. \cite{hasan}. Solid curves denote quark model predictions with boosted intrinsic
         pion wave functions; dashed lines represent predictions with original (spherical) pion wave functions.}
\label{fig1}
\end{figure*}
for the {\em transversal\/} component. The factor $\mathcal{N}$ is an overall normalization and $\mathbf{R'}=\mathbf{R}_{\pi^-}\!\!-\mathbf{R}_{\pi^+}$ 
and $\mathbf{R}=\mathbf{R}_{\bar p}-\mathbf{R}_p$ are the relative $\pi^-\pi^+$ and antiproton-proton coordinates, respectively.
The angle $\theta$ is between $\mathbf{R'}$ and $\mathbf{R}$ in the c.m. frame. In the experiment it is the c.m. angle between the
antiproton beam direction and the outgoing $\pi^-$ direction. The new terms mentioned previously are $C_V$, $C_L$, $C_T$, and $D$.
The selection rules for the $R2$ diagrams discussed in Ref.~\cite{kloet1} are unchanged. More precisely, $\hat T(\p)$ and $\hat T(\s^L)$ 
act in $\bar pp$ states with $J^\pi=0^+, 2^+, 4^+,...$  waves while $\hat T(\s^T)$ contributes only to $J^\pi=1^-, 3^-, 5^-,...$ waves. 
The explicit expressions for the coefficients of Eqs.~(\ref{tot1})--(\ref{tot3}) are (with $\alpha$ being the proton size parameter defined 
similarly to $\beta$ in Eq.~(\ref{pionnoboost})):
\begin{eqnarray}
 A & = & -\frac{\alpha(5\alpha+4\beta\gamma^2)}{2(4\alpha+3\beta\gamma^2)} \\
 B & = & -\frac{3(7\alpha^2 +18\alpha\beta\gamma^2+9\beta^2\gamma^4)}{8(4\alpha+3\beta\gamma^2)}-D \\
 C & = & -\frac{3\alpha(\alpha+\beta\gamma^2)}{2(4\alpha+3\beta\gamma^2)} \\ 
 D & = & -\frac{9\beta(\gamma^2-1)}{8}\!\left \{ 1+\frac{\alpha^2}{(5\alpha+4\beta)(5\alpha+4\beta\gamma^2)}\right\}
\hspace{8mm}
\end{eqnarray}
\begin{eqnarray}
 A_V & = & \frac{\alpha(\alpha+\beta\gamma^2)}{4\alpha+3\beta\gamma^2}  \\
 B_V & = & \frac{3(\alpha+\beta\gamma^2)(5\alpha+3\beta\gamma^2)}{2(4\alpha+3\beta\gamma^2)} - C_V  \\
 C_V & = & \frac{3\beta(\gamma^2-1)}{2}\left \{ 1+\frac{\alpha^2}{(5\alpha+4\beta)(5\alpha+4\beta\gamma^2)} \right \} 
\hspace{8mm}
\end{eqnarray}
\begin{eqnarray}
 A_L & = & -A_V  \\
 B_L & = & \frac{9(\alpha+\beta\gamma^2)^2}{2(4\alpha+3\beta\gamma^2)}-C_L \\
 C_L & = & \frac{3\beta(\gamma^2-1)}{2}\left \{ 1-\frac{\alpha^2}{(5\alpha+4\beta)(5\alpha+4\beta\gamma^2)} \right \}
\hspace{8mm}
\end{eqnarray}  
\begin{eqnarray}
 A_T & = & -2A_V \\
 B_T & = &  \frac{3\alpha(\alpha+\beta\gamma^2)}{4\alpha+3\beta\gamma^2}-C_T \\
 C_T & = & - \frac{3\beta\alpha^2(\gamma^2-1)}{(5\alpha+4\beta)(5\alpha+4\beta\gamma^2)} .
\hspace{3.4cm}
\end{eqnarray} 
\begin{figure*}[t]
\begin{center}
\includegraphics[angle=90,scale=0.55]{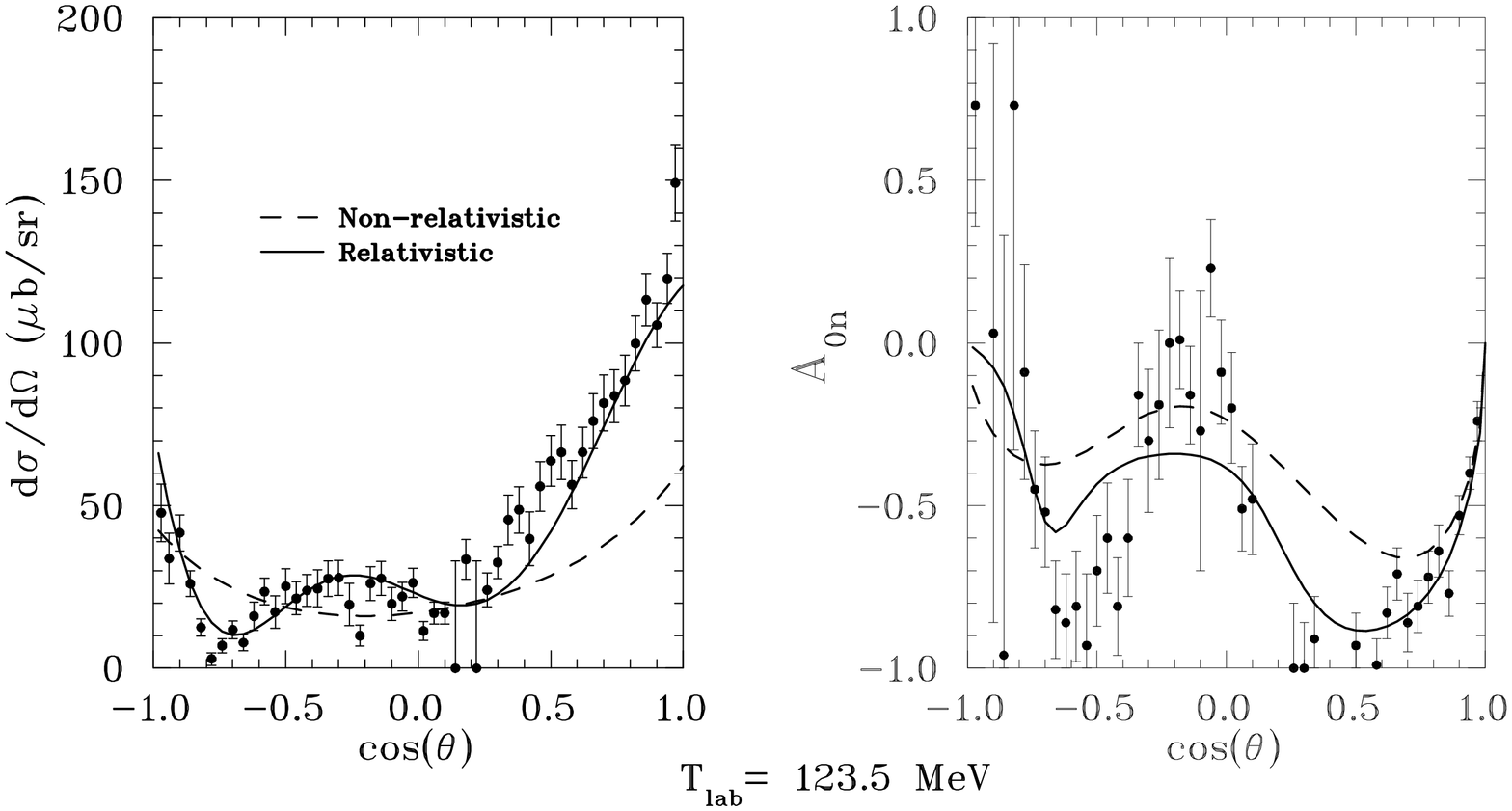}
\end{center}
\caption{\small As in Fig.~\ref{fig1} but for $\tlab$= 123.5 MeV  ($\plab$ = 497 MeV/c).}
\label{fig2}
\begin{center}
\includegraphics[angle=90,scale=0.55]{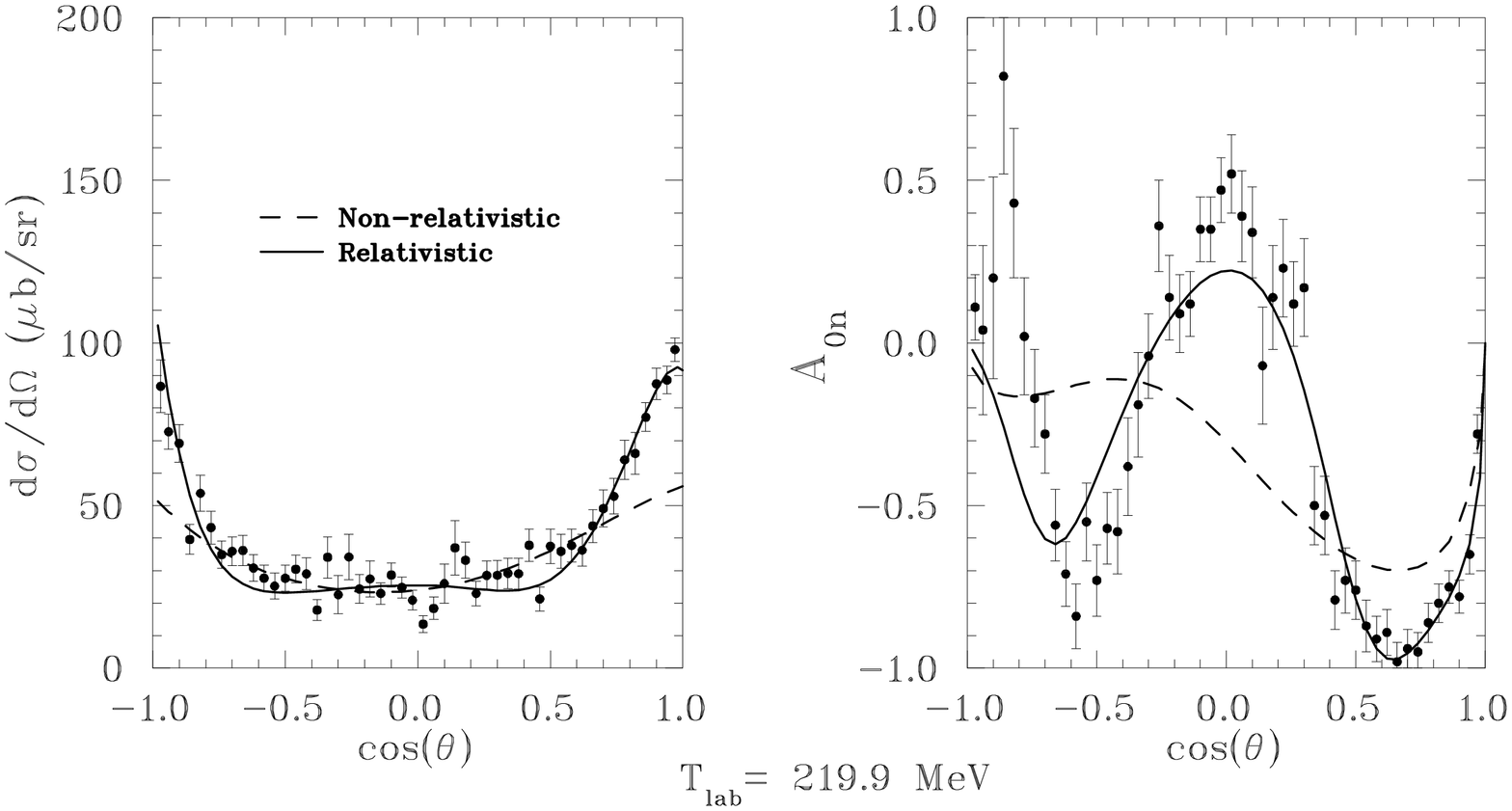}
\end{center}
\caption{\small As in Fig.~\ref{fig1} but for  $\tlab$= 219.9 MeV ($\plab$= 679 MeV/c).}
\label{fig3}
\end{figure*}
Clearly, $C_V$, $C_L$, $C_T$, and $D$ vanish in the non-relativistic limit $\gamma\rightarrow 1$. A noticeable feature is that some of the 
new coefficients become relatively large for large $\gamma$ values. In particular, the $D R^2 \cos^2\theta$ term dominates the exponential part of 
Eqs.~(\ref{tot1})--(\ref{tot3}) and, for example, $C_V\gg A_V, B_V$ and $C_L\gg A_L, B_L$. On the other hand, one finds that $C_T \simeq A_T \simeq B_T$.

At this stage we will neglect additional relativistic corrections due to the (anti)quark spinors in the final state pions since it was argued in 
Ref.~\cite{elbennich2} that boosting of spin wave functions results in relatively smaller corrections.

\begin{figure*}[t]
\begin{center}
\includegraphics[angle=90,scale=0.55]{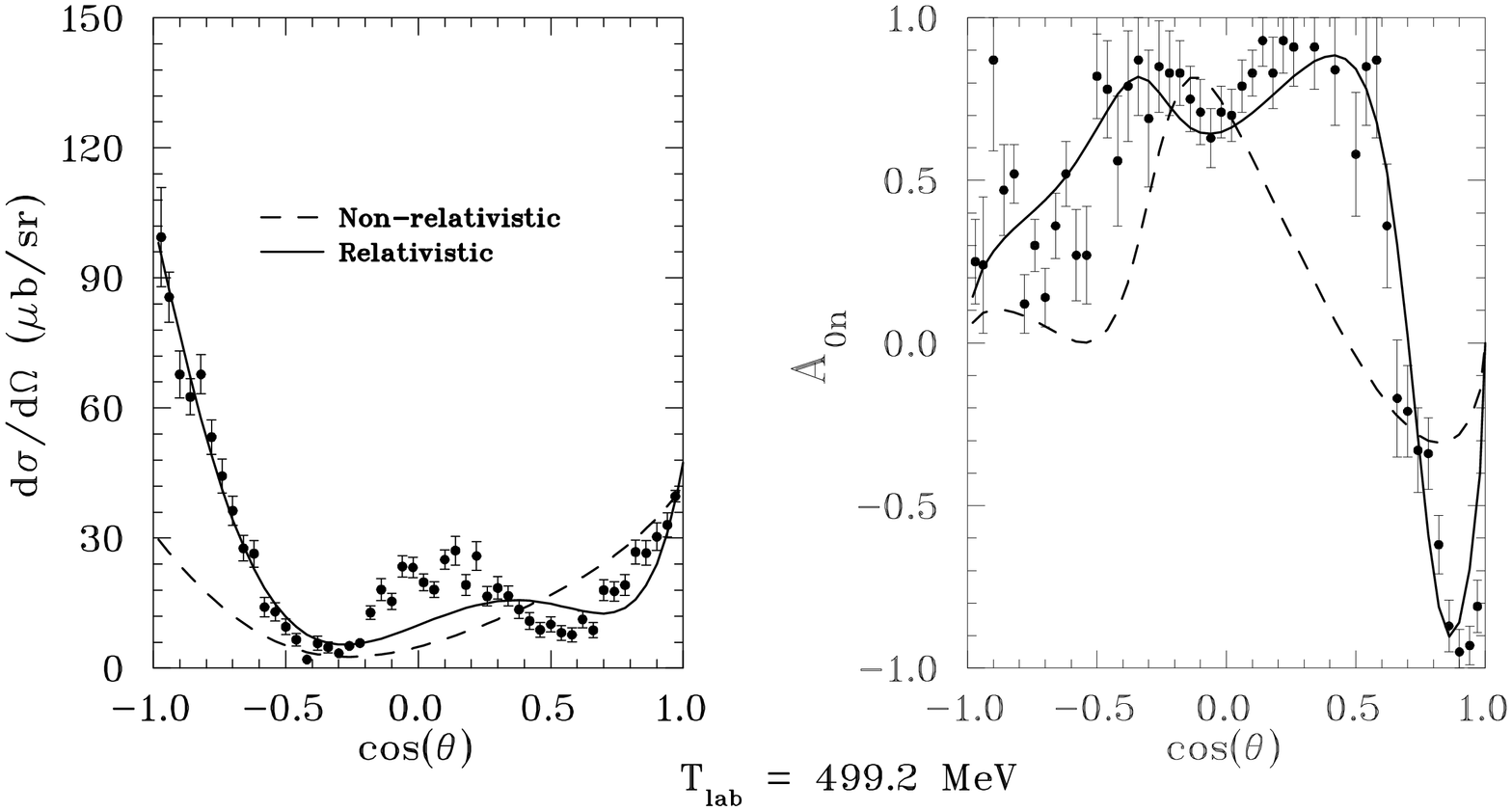}
\end{center}
\caption{\small As in Fig.~\ref{fig1} but for $\tlab$= 499.2 MeV ($\plab$= 1089 MeV/c).}
\label{fig4}
\begin{center}
\includegraphics[angle=90,scale=0.55]{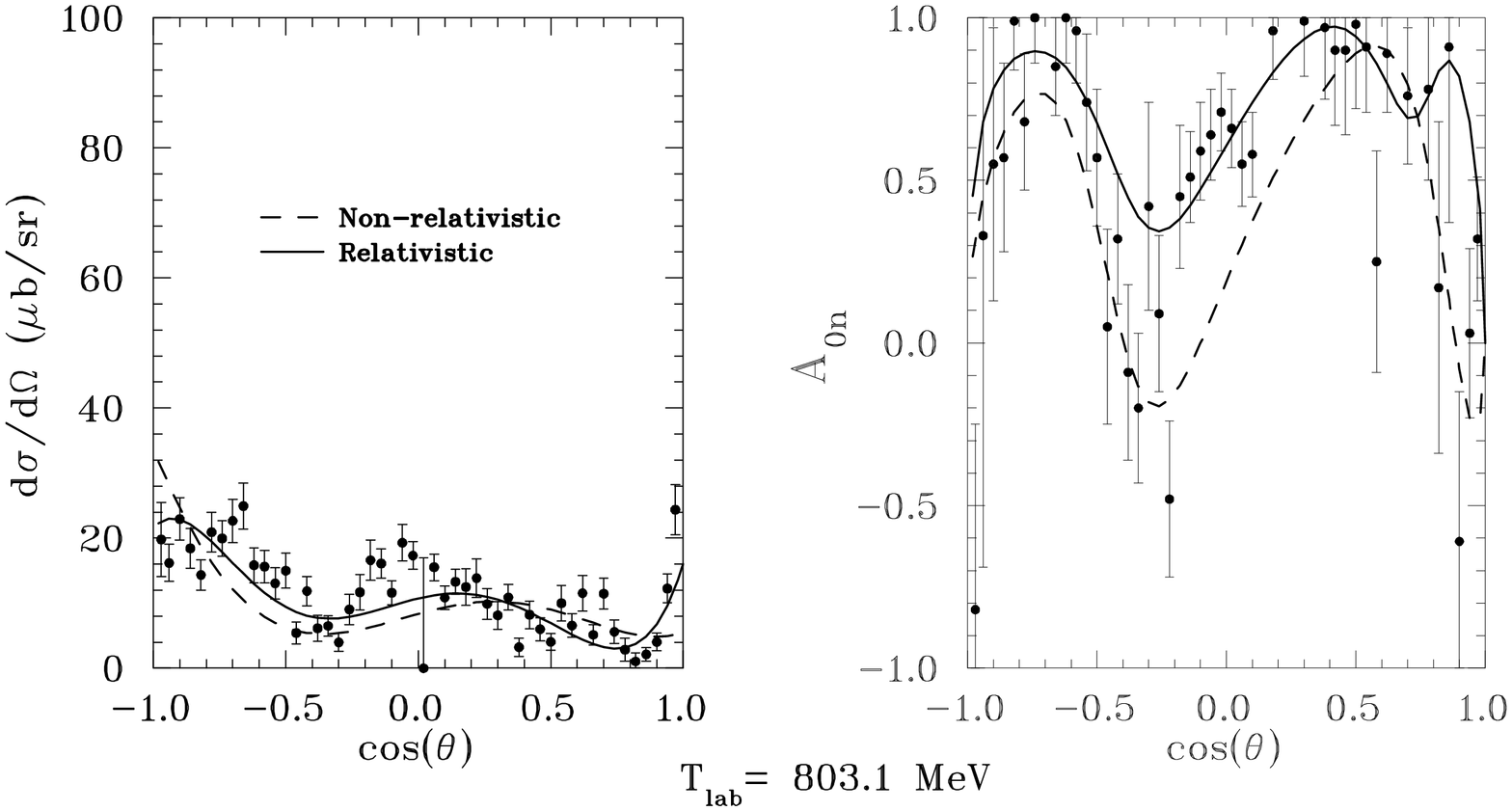}
\end{center}
\caption{\small As in Fig.~\ref{fig1} but for $\tlab$= 803.1 MeV  ($\plab$= 1467 MeV/c).}
\label{fig5}
\end{figure*}

\section{Results}

The LEAR data for the $d\sigma/d\Omega$ and $A_{0n}$ have been fitted using the transition operators in Eqs.~(\ref{tot1}), (\ref{tot2}), 
and (\ref{tot3}). As before, taking only the $R2$ topology into account, we use a distorted wave approximation (DWA) 
\begin{equation}
 T^J =\int\! \mathrm{d}\mathbf{R} \mathrm{d}\mathbf{R'}\, \phi_{\pi\pi}^J(\mathbf{R'})\,
       \hat T_{R2} (\mathbf{R},\mathbf{R'})\,\Psi_{\bar pp}^{J, l=J\pm 1}(\mathbf{R}) 
\label{int}
\end{equation}
to obtain the scattering matrix elements and introduce final-state interactions as in Ref.~\cite{elbennich1}. The pion wave function 
$\phi_{\pi\pi}^J(\mathbf{R'})$ is obtained from the coupled-channel model of Ref.~\cite{kloet3} and can be parameterized with phases 
$\delta_J$ and inelasticities $\eta_J$. The initial $\bar pp$ wave function $\Psi_{\bar pp}^{J, l=J\pm 1}(\mathbf{R})$ is taken from 
an optical potential model~\cite{elbennich3}. Both  $\phi_{\pi\pi}^J(\mathbf{R'})$ and $\Psi_{\bar pp}^{J, l=J\pm 1}(\mathbf{R})$ still 
contain the angle dependence appropriate for the values of $J$ and $l$.

We select from the specific energies where LEAR data are available \cite{hasan} the same set of five energies as in Ref.~\cite{elbennich1}: 
$\tlab=66.7$, 123.5, 219.9, 499.2, and 803.1 MeV corresponding to antiproton momenta of respectively $\plab= 360$, 497, 679, 1089, 
and 1467 MeV/$c$. $\tlab$ is the laboratory kinetic energy of the antiproton beam. In fitting the data, we proceed as in Ref.~\cite{elbennich1}. 
One starts with $\pi^-\pi^+$ final-state plane waves, i.e. with $\delta_J=0$ and $\eta_J=1$, and varies these parameters for $0\le J\le 4$. 
The other parameters are the relative (complex) strength $\lambda$ as defined by the $R2$ transition amplitude
\begin{equation}
\hspace*{-1mm} \hat  T_{\mbox{\tiny tot.}} ({\mathbf R}'\!,{\mathbf R})= 
  N_0\, [ \hat T(\p)({\mathbf R}'\!,{\mathbf R}) + \lambda\, \hat T(\s)({\mathbf R}'\!,{\mathbf R}) ] ,
\end{equation}
and the overall normalization $N_0$. The size parameters $\alpha$ and $\beta$ of the quark model are kept fixed at the original values 
$\alpha=2.8~$fm$^{-2}$ and $\beta=3.23~$fm$^{-2}$ \cite{kloet1,kloet2,green}. The aim is to find out how much improvement is obtained solely
from introducing final-state interactions. 

At this initial stage, no relativistic effects have been included which implies $\gamma=1$. The results of this fit are plotted in 
Figs.~\ref{fig1}--\ref{fig5} as dashed lines. They will serve as reference point and were obtained previously in  Ref.~\cite{elbennich1} 
where the model with and without final-state interactions was discussed. It was found that the improvement from just final-state interactions 
is very modest --- the double-dip structure in the analyzing power is somewhat more enhanced but the forward peaks in the differential cross
sections at $\tlab= 66.7$ and 123.5 MeV ($\plab= 360$ and  497 MeV/$c$) as well as the backward peaks at $\tlab=499.2$ and 803.1 MeV 
($\plab=1089$ and 1467 MeV/$c$) are poorly reproduced.

Next, we consider the effects of relativistic distortions of the intrinsic pion wave functions. Hence, the transition operators of 
Eqs.~(\ref{tot1})--(\ref{tot3}) must be used. The exponential part is dominated by the $D$ term (typically $D \simeq 200~$fm$^{-2}$, about 
two orders of magnitude larger than the $C$ term), which makes the partial wave amplitudes $T^J$ very sensitive to the value of the relative 
strength $\lambda=|\lambda|\exp(i\theta_{\lambda})$ of the $\s$ versus the $\p$ mechanism and to the pion wave parameters $\eta_J$ and 
$\delta_J$. This is a consequence of the strong angle dependence of $\exp\{ DR^2\cos^2\theta\}$. In order for the partial wave expansion 
of the scattering amplitude $T$ to converge, all ingredients must be fine-tuned, in particular the phases $\delta_J$ and inelasticities 
$\eta_J$ for $0\le J\le 4$. It should be emphasized that $\alpha$ and $\beta$ remain fixed.

The improved new fit is illustrated in Figs.~\ref{fig1}--\ref{fig5}, where the predictions that include relativistic effects
and final-state interactions are presented as solid curves. The new fit is a significant improvement over the non-relativistic version of the
$R2$ annihilation model. The differential cross sections $d\sigma/d\Omega$ are particularly well reproduced for the energies $\tlab= 66.7$, 
123.5, and  219.9~MeV ($\plab = 360$, 497, and 679~MeV/$c$). For $\tlab=499.2$ and 803.1~MeV ($\plab = 1089$ and 1467~MeV/$c$) the results for 
$d\sigma/d\Omega$ show improvement in forward and backward direction. At the two higher energies, the experimental double-well structure of the
cross section is reproduced but some problems remain near $\theta=\pi/2$.

As for the analyzing powers $A_{0n}$, we are able to reproduce their double-dip structure, already present in the LEAR data at the lowest energies.
When the experimental errors are small, like for instance for $\tlab= 219.9$ and 499.2~MeV, the fit is particularly successful. At lowest energy, 
$\tlab=66.7$~MeV, the error bars in $A_{0n}$ are rather large and we do not fit very well certain data points. Overall the improvement is 
clearly important and of similar quality as the results of Ref.~\cite{elbennich1} which were obtained from increasing the particle radii.

\begin{table}[t]
\begin{ruledtabular}
  \begin{tabular}{c|rrrrr}   
    $\tlab$ [MeV]  & 66.7  & 123.5 & 219.9 & 499.2 & 803.1 \\
    $\plab$ [MeV/$c$] & 369 & 497 & 679 & 1089 & 1467  \\ 
    $\gamma=\sqrt{s}/2m_{\pi}$ & 6.841 & 6.940 & 7.106 & 7.564 & 8.033 \\  \hline
    $|\lambda|$ & 1.165 & 0.889 & 1.090 & 1.377 & 1.320 \\
    $\theta_{\lambda} [^{\circ}]$ & 154.51 & 149.13 & 82.57 & 184.14 & 184.76 \\ 
    $N_0$ $[10^5]$ & 2.999 & 3.061 & 3.068 & 1.739 & 4.999 
  \end{tabular}
\end{ruledtabular}
\caption{\small Quark model parameters as function of $\tlab$.}
\label{tab1}
\end{table}

\begin{table}
\begin{ruledtabular}
  \begin{tabular}{c|rrrrr}
    $\tlab$ [MeV]  &  66.7  & 123.5 & 219.9 & 499.2 & 803.1 \\
    $\sqrt s$ [MeV] & 1910. & 1937. & 1983. & 2111. & 2242. \\ \hline
    $\eta_0$ & 1.00 & 0.716 & 0.811 & 1.00 & 1.00 \\
    $\delta_0$ & 67.00 & 54.37 & 41.42 & $-1.704$ & $-17.15$ \\ 
    $\eta_1$ & 1.00 & 0.583 & 0.666 & 0.999 & 1.00 \\
    $\delta_1$ & 43.68 & 37.11 & 42.62 & $-31.62$ & 21.56 \\
    $\eta_2$ & 0.968 & 0.973 & 0.992 & 0.538 & 0.624 \\
    $\delta_2$ & $-17.06$ & $-16.55$ & $-18.78$ & $-37.34$ & $-13.51$ \\ 
    $\eta_3$ & 1.00 & 0.923 & 1.00 & 0.946 & 1.00 \\ 
    $\delta_3$ & $-11.93$ & $-13.80$ & $-16.07$ & $-6.873$ & $-11.91$ \\
    $\eta_4$ & 1.00 & 1.00 & 1.00 & 0.946 & 1.00 \\
    $\delta_4$ & $-0.03$ & $-0.03$ & $-0.04$ & $-0.038$ & $-0.07$ 
  \end{tabular}
\end{ruledtabular}
\caption{\small Phases shifts $\delta_J$ and inelasticities $\eta_J$ of the final-state $\pi\pi$ interaction for $0\le J\le 4$ as function of $\sqrt s$.}
\label{tab2}
\end{table}

In Tab.~\ref{tab1} we give the values of the quark model parameters as a function of $\tlab$ while the $\pi\pi$ phases $\delta_J$ and inelasticities 
$\eta_J$ are conveniently listed as a function of c.m. energy $\sqrt{s}$ in Tab.~\ref{tab2}. In the quark model parameters of Tab.~\ref{tab1} one 
notes that the relative strength $|\lambda|$ of the $\s$ versus the $\p$ mechanism varies with energy and the average value $|\lambda|\simeq 1.17$. 
There is no {\em a priori\/} reason why $\lambda$ should be energy independent. The energy variation of the inelasticities $\eta_J$ in the $\pi\pi$ 
interaction observed in Tab.~\ref{tab2} may be associated with the presence of $\pi\pi$ resonances, in particular with the $f_2$ at $2010$ MeV in 
the $J=2$ amplitude.

\section{Summary and Conclusion}

We present a new fit to the LEAR $\pppi$ data based on relativistic effects in the annihilation. These effects are due to Lorentz transformations 
of intrinsic pion wave functions from the pion rest frame to the c.m. frame of the reaction. The impact from the modified geometry of the pions is 
considerable and carries with it a much richer angle dependence of the transition operators.

The quality of this new fit is very similar to the one of a previous fit \cite{elbennich1}. However, the approaches are very different.
While in Ref.~\cite{elbennich1} we achieved a good fit by means of particle radii increase, so the antiquarks and quarks do overlap
more, in this work we keep the original particle sizes fixed but take into account the relativistic distortions of the pions in the c.m.
frame of the reaction.

We conclude that even though observed data at LEAR on $d\sigma/d\Omega$ and $A_{0n}$ indicate strongly the presence of higher partial wave 
amplitudes and therefore seem to call for an increase of the annihilation range, one can obtain a similar improvement by additional angle 
dependence in the transition operators. This additional angle dependence arises naturally when boosted intrinsic pion wave functions are 
employed in the quark model calculations, as becomes clear from the present fit. Relativistic effects in the reaction $\pppi$ therefore should 
not be ignored.

\end{document}